# Visual Accessibility in a Virtual Kitchen: Effects of Open Shelving on Performance, Cognitive Load, and Experience in Older Adults with and without MCI

Ibrahim Bilau, Eunhwa Yang, Hyeokhyen Kwon, Stacie Smith, Bruce Walker, Hui Cai, Ece Erdogmus, Omobolanle Ogunseiju

**Abstract**

This study examines how visual accessibility through cabinet design influences task performance, cognitive load, physical activity level, motivation, and user experience in a virtual kitchen among older adults with and without mild cognitive impairment (MCI). Seventeen older adults (7 with MCI, 10 without) completed a repeated-measures item-retrieval task under two conditions, closed cabinets and open shelving, using a counterbalanced within-subjects design. Measures included task duration, physical activity level (ENMO), cognitive load (NASA-TLX and gaze entropy), intrinsic motivation (IMI), and post-task interviews. Open shelving significantly reduced task duration (β = −291.20, p < .001) and physical activity level (β = −0.00615, p = .008). Gaze entropy increased (β = 1.29, p = .001), with a significant Setting × MCI interaction (p = .009) and moderation by MoCA score (p < .001). NASA-TLX and intrinsic motivation did not differ significantly between conditions. Qualitative findings indicated reduced reliance on memory-based search and highlighted themes related to independence, aesthetics, safety, and adoption. Overall, visual accessibility improved efficiency and reduced movement demands while altering visual-search organization, with divergence between subjective and objective indicators of cognitive load. These findings support visually accessible design strategies to enhance functional performance and inform cognitively supportive built environments for aging populations.

# Introduction

Aging in place, defined as the ability of older adults to remain in their own homes and communities as they age, is widely recognized as a critical goal for individuals and health systems (Wiles et al., 2012). However, age-related cognitive and physical changes can make everyday activities more difficult, particularly tasks that require memory, planning, visual search, and coordinated action (Lawton & Brody, 1969). Mild cognitive impairment (MCI) represents an intermediate stage between normal aging and dementia and is associated with declines in memory and executive functioning that can interfere with everyday activities (Alzheimer's Association, n.d.; Manly et al., 2022; Petersen et al., 2009). The prevalence of MCI has been estimated at approximately 12% to 18% among adults aged 60 years and older (Alzheimer's Association, n.d.; Manly et al., 2022; Ward et al., 2012). These challenges are especially evident in instrumental activities of daily living (IADLs), which require cognitive and physical coordination (Perneczky et al., 2006; Teng et al., 2010).

Kitchen environments are central to IADLs such as meal preparation, food storage, and household organization (Lawton & Brody, 1969; Wiles et al., 2012). Empirical work shows that older adults experience difficulty retrieving and organizing items in kitchen environments, with poor design increasing physical strain and task difficulty (Iwarsson et al., 2007; Zhou et al., 2024). Research on kitchen ergonomics has identified visibility, reachability, and organization as recurring design challenges (Maguire et al., 2014; Hrovatin et al., 2015). More recent work demonstrates that home design factors such as spatial complexity, visual clutter, storage visibility, and accessibility directly influence IADL performance among older adults with MCI (Machry et al., 2025).

Visual accessibility has been proposed as a design strategy to reduce these demands. Open shelving allows users to identify stored objects directly without opening cabinet doors, which may reduce search time and unnecessary movement while supporting more efficient task execution (Fleming et al., 2016; Norman, 2013). Increasing the visibility of task-relevant objects may be particularly beneficial for individuals with MCI because it reduces reliance on memory during task performance (Gitlin, 2003; Nygard, 2004). Prior work suggests that visually accessible kitchen storage can improve task efficiency, reduce physical effort, and enhance user experience, although responses may vary due to concerns such as clutter and aesthetics (Bilau et al., 2025).

The rationale for visually accessible storage is consistent with established theory. Cognitive Load Theory (CLT) distinguishes between intrinsic task demands and extraneous demands introduced by the way information is presented (Sweller, 1988; Paas et al., 2003). Concealing objects behind cabinet doors may increase extraneous cognitive load by requiring memory-dependent and iterative search, whereas open shelving may reduce that burden by providing direct visual access (Chandler & Sweller, 1991; Young et al., 2015). Environmental Press Theory (EPT) suggests that functional outcomes depend on the alignment between an individual's capabilities and environmental demands (Lawton & Nahemow, 1973).

Studying these design effects in physical environments presents methodological challenges. Real-world kitchen studies are difficult to standardize due to variability in layouts, storage configurations, and object placement (Hrovatin et al., 2015; Maguire et al., 2014). Traditional assessments of everyday function often rely on self-report or informant-based measures, which may be subject to bias or inaccuracy (Kaplan et al., 2026). These limitations make it difficult to isolate the effects of specific environmental features and to capture detailed behavioral responses.

Virtual reality (VR) enables controlled manipulation of environmental features within immersive, spatially structured environments (Parsons, 2015; Parsons et al., 2017; Rizzo & Koenig, 2017). VR-based assessments can differentiate functional performance between cognitively healthy older adults and those with MCI in simulated tasks (Kaplan et al., 2026). VR systems can also be integrated with sensing technologies, enabling simultaneous measurement of visual attention, movement patterns, and task performance (Parsons, 2015; Rizzo & Koenig, 2017).

The present study examines the effects of cabinet visibility in a virtual kitchen environment modeled from a real-world setting. Older adults with and without MCI completed a standardized item-retrieval task under two conditions: closed cabinets and open shelving. A repeated-measures design was used to isolate the effect of cabinet configuration. Outcome measures included task duration, physical activity level derived from wrist accelerometry, cognitive load assessed through subjective (NASA-TLX) and objective (gaze entropy) measures, and intrinsic motivation (IMI). Semi-structured interviews captured participants' perceptions of usability, effort, and independence.

Based on these frameworks, the study addresses the following research questions:

**RQ1.** How does visual accessibility through open shelving influence cognitive load, intrinsic motivation, physical activity level, and task duration compared to closed cabinets during a virtual kitchen retrieval task?

**RQ2.** How do task performance and cognitive load outcomes differ between older adults with MCI and those without MCI during the virtual kitchen retrieval task?

**RQ3.** How does cognitive status (MCI versus non-MCI) moderate the effects of cabinet visibility on task outcomes?

**RQ4.** How does continuous cognitive status, as measured by the MoCA score, influence the relationship between cabinet visibility and task outcomes?

**RQ5.** To what extent do subjective workload and effort reports converge with or diverge from objective task and visual search measures?

## Methods

### Study Design

This study used a mixed-methods repeated-measures design to evaluate the effects of cabinet visibility on functional performance during a standardized item-retrieval task. The independent variable was cabinet visibility, operationalized as Open shelving versus Closed cabinets. Each participant completed the task under both conditions, creating a within-subjects repeated-measures design, while cognitive status served as a between-subjects factor. A convergent mixed-methods approach was used to analyze and integrate quantitative and qualitative strands (Creswell & Plano Clark, 2018; Fetters et al., 2013; Guetterman et al., 2015).

**Participants**

Participants were recruited through community-linked aging networks in the Atlanta metropolitan area. Eight participants were recruited as MCI–care-partner dyads through a structured program; remaining participants were recruited through general community outreach. The final sample included 17 older adults (see Table 1): 7 with MCI (mean age = 77.0 years, SD = 6.3; 4 male, 3 female; 4 Black, 3 White) and 10 without MCI (mean age = 74.1 years, SD = 5.7; 1 male, 9 female; 4 Black, 6 White). MoCA scores ranged from 20 to 22 (M = 21.7, SD = 0.8) in the MCI group and from 26 to 29 (M = 28.2, SD = 1.1) in the non-MCI group. Data were structured in long format, yielding 34 participant-condition observations. Four CEP dyad-linked participants contributed to the qualitative interview data. This research complied with the American Psychological Association Code of Ethics and was approved by the Institutional Review Board at Georgia Institute of Technology. Informed consent was obtained from each participant.

*Table 1. Participant characteristics, including age, gender, MoCA score, cognitive status, dyad ID, and race (n = 17)*

| ID | Age | Gender | MoCA | Cognitive Status | Dyad ID | Race |
|---|---|---|---|---|---|---|
| 0001 | 83 | F | 29 | Non-MCI | 0001 | W |
| 0002 | 88 | F | 25 | MCI | 0002 | W |
| 0003 | 67 | F | 26 | Non-MCI | 0003 | B |
| 0004 | 85 | F | 28 | Non-MCI | 0004 | W |
| 0005 | 82 | M | 20 | MCI | 0005 | W |
| 0006 | 83 | F | 29 | Non-MCI | 0005CP | W |
| 0007 | 78 | F | 27 | Non-MCI | 0006 | B |
| 0008 | 72 | F | 29 | Non-MCI | 0007 | B |
| 0009 | 63 | M | 21 | MCI | 0008 | B |
| 0010 | 76 | F | 29 | Non-MCI | 0008CP | B |
| 0011 | 78 | M | 21 | MCI | 0009 | B |
| 0012 | 80 | F | 25 | MCI | 0009CP | B |
| 0013 | 72 | F | 27 | Non-MCI | 0010 | W |
| 0014 | 81 | M | 22 | MCI | 0011 | B |
| 0015 | 72 | F | 29 | Non-MCI | 0012 | W |
| 0016 | 70 | M | 22 | MCI | 0013 | B |
| 0017 | 71 | F | 29 | Non-MCI | 0013CP | W |

*Note. F = Female; M = Male; W = White; B = Black; CP = Care Partner. Care-partner participants were recruited through a structured program alongside their paired MCI participants, identified by a shared Dyad ID (e.g., 0005 and 0005CP).*

All procedures were conducted under approval from the college's institutional review board. Participants provided informed consent and could withdraw at any time. Eligibility required ability to complete study procedures in English and engage with the VR task. Cognitive status was based on existing records or assessed using the Montreal Cognitive Assessment (MoCA). Exclusion criteria included inability to consent, mobility limitations, discomfort with the VR headset, or technical issues. Missing values were handled using available-case analysis.

**Apparatus and Materials**

Experimental sessions used an HTC VIVE Pro head-mounted display (2880 × 1600 resolution, 90 Hz) with SteamVR Tracking 2.0. Participants interacted using hand controllers. Calibration and tracking verification were completed before each session.

Three synchronized data sources were collected: survey data (participant identifiers, cognitive group, order variables, and self-report outcomes), wrist accelerometry from GeneActiv sensors, and eye-tracking data from the Pupil Labs VR system. Sensor features were aligned to common task intervals.

Self-report measures included the Simulator Sickness Questionnaire (SSQ; Kennedy et al., 1993), the Positive and Negative Affect Schedule (PANAS; Watson et al., 1988), the NASA Task Load Index (NASA-TLX; Hart & Staveland, 1988), the Intrinsic Motivation Inventory (IMI; Ryan, 1982), the Borg Rating of Perceived Exertion scale (Borg, 1982), and the System Usability Scale (SUS; Brooke, 1996). The analysis focused on task duration, physical activity level, cognitive load, intrinsic motivation, and qualitative perceptions.

**Virtual Environment**

The virtual kitchen was developed as a three-dimensional reconstruction of the first-floor kitchen in a living lab (Kidd et al., 1999). The environment reflected a typical residential layout with an L-shaped counter and realistic lighting.

The physical kitchen was digitized using a Faro Focus S70 scanner, processed in Faro SCENE, retopologized in SketchUp, and implemented in Unreal Engine 5 using Datasmith and Lumen. The final scene was optimized for real-time rendering (Bilau et al., 2026).

The primary manipulation was cabinet visibility. In the Closed condition, participants opened cabinet doors to locate items. In the Open condition, cabinet contents were visible without door interaction. All other elements (layout, object locations, lighting, and task workflow) were held constant. The two conditions are illustrated in Figure 1.

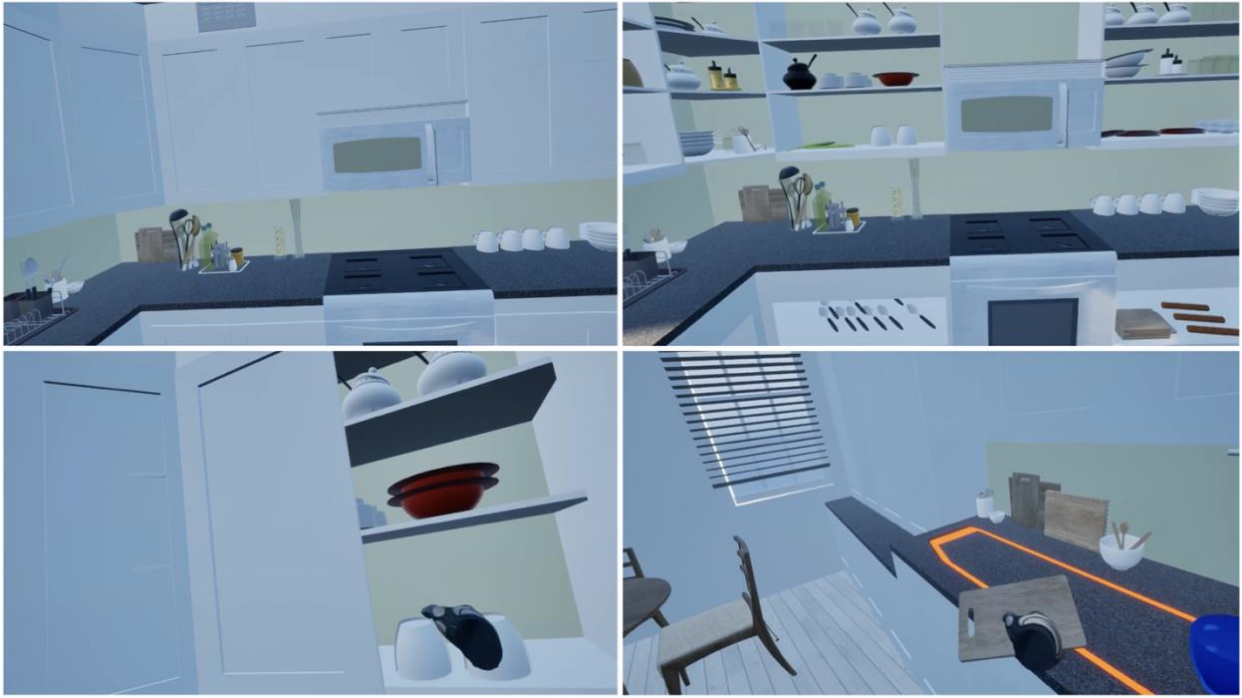

*Figure 1.* Virtual kitchen environment showing the two cabinet visibility conditions. Upper left panel: Closed cabinet condition, in which all cabinet doors are shut, and participants must open them to locate items. Upper right panel: Open shelving condition, in which cabinet contents are fully visible without door interaction.

**Task and Procedure**

Participants completed a VR item-retrieval task, locating and placing objects in a designated counter area. The object sequence was standardized across participants and conditions. Limited assistance was provided when necessary. Participants interacted with objects in the environment by using the VIVE hand controllers: they pointed at target objects with the controller ray and pressed the trigger button to grasp and move items to the designated counter area. A brief familiarization trial was completed prior to the task to ensure controller proficiency. Prior VR experience was minimal; only two participants reported any previous VR use, and both required the full familiarization trial.

Condition order was counterbalanced using block randomization (Efird, 2011; see Figure 2 for the full session sequence). Participants completed baseline measures (SSQ, PANAS), followed by headset calibration and familiarization. After a rest period, participants completed Condition 1, followed by post-condition surveys (SSQ, PANAS, NASA-TLX, IMI, Borg). Participants then completed Condition 2 and a final survey block, followed by a semi-structured interview. Breaks were included to reduce fatigue.

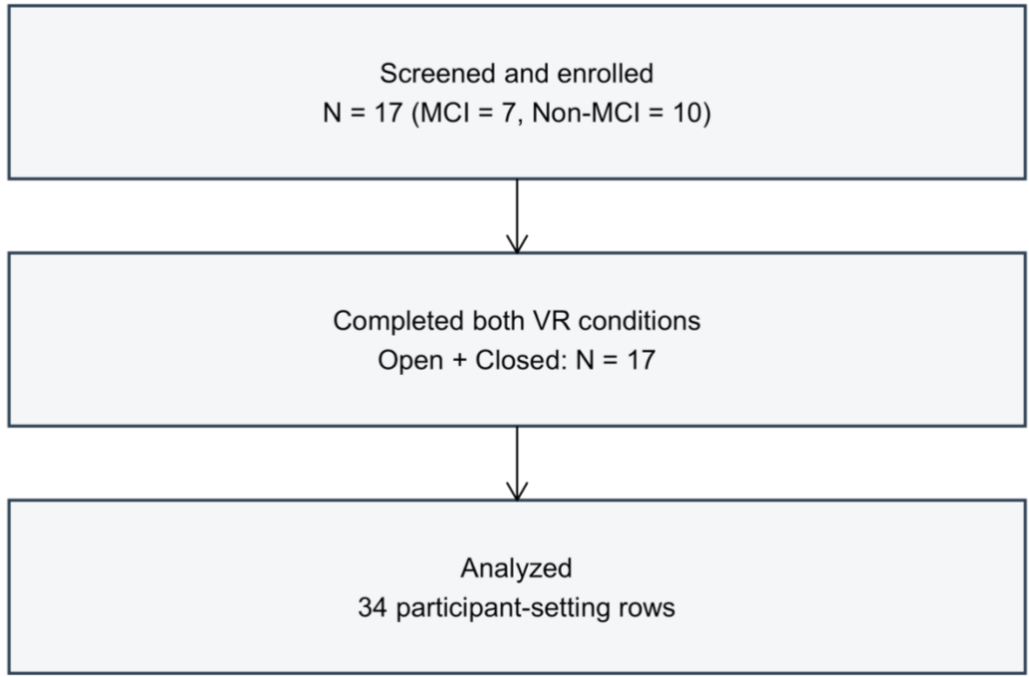

*Figure 2*. Experimental session flow showing the sequence of activities completed by each participant. The session included pre-session consent and screening, VR setup and calibration, a familiarization trial, two counterbalanced task conditions (Open and Closed) with rest breaks and post-condition surveys between them, and a concluding semi-structured interview. Condition order was block-randomized across participants.

*Table 2*. Experimental session sequence for the study.

| Session Phase | Activity | Details |
| --- | --- | --- |
| Pre-session | Consent and screening | Informed consent, MoCA administration, SSQ and PANAS administered |
| Setup | Equipment fitting | HTC VIVE Pro HMD fitted and calibrated; room and floor alignment completed |
| Familiarization | VR orientation | Brief introduction to VR environment and controller interaction |
| Break | Rest | Minimum two-minute break |
| Condition 1 | Item retrieval task | Participant completes retrieval task in first assigned condition (Open or Closed, randomized) |
| Break | Rest and survey | Minimum five-minute break: SSQ, PANAS, NASA-TLX, IMI and Borg scale surveys completed |
| Condition 2 | Item retrieval task | Participant completes retrieval task in second condition |
| Post-session | Final surveys | SSQ, PANAS, NASA-TLX, IMI and Borg scale surveys completed; semi-structured interview conducted |

## Measures

### Task Performance

Task duration represented the time required to complete the retrieval task under each condition.

### Physical Activity Level

Physical activity level was derived from wrist accelerometry using the Euclidean Norm Minus One (ENMO; see Equation 1):

$$\boldsymbol{ENMO = max\left(\sqrt{X^2 + Y^2 + Z^2} - 1, 0\right)}$$

*Equation 1.* ENMO computed from tri-axial accelerometer signals (X, Y, Z), with negative values truncated to zero.

Perceived physical exertion was assessed using the Borg scale, providing a subjective complement to the objective accelerometry measure.

### Cognitive Load

Subjective workload was measured using NASA-TLX. Objective indicators were derived from eye-tracking, with gaze entropy as the primary outcome. Additional exploratory features included fixation duration, pupil variability, and valid-gaze percentage.

### Motivation

Intrinsic motivation was assessed using IMI subscales (interest/enjoyment, perceived competence, pressure/tension).

### Qualitative Perceptions

After completion of both VR conditions, all 17 participants completed semi-structured interviews addressing usability, accessibility, independence, safety, and adoption.

## Data Analysis

### Quantitative Analysis

Data were processed to ensure alignment across survey, accelerometry, and eye-tracking streams using survey records, task timing records, and video timestamps. Invalid eye samples were excluded. One participant with missing eye-tracking data for one condition was excluded from eye-based analyses. Across remaining files, 86.64% of samples were retained.

Primary analyses used linear mixed-effects models (lme4; lmerTest, Satterthwaite approximation; Bates et al., 2015; Kuznetsova et al., 2017) specified as follows (Equation 2):

$$DV \sim Setting \times MCI + Start\_Order + (1 | Part\_ID)$$

**Equation 2**. *Linear mixed-effects model specification including Setting × MCI interaction, Start Order, and a random intercept for participant (Part_ID).*

Closed cabinet condition and non-MCI were reference levels. Holm correction (Holm, 1979) was applied to primary endpoints (task duration, ENMO mean, NASA-TLX, gaze entropy). Exploratory analyses used MoCA as a continuous moderator. Spearman correlations examined subjective–objective associations. Sensitivity analyses included leave-one-out and random-effects checks.

A secondary model included dyad clustering (Equation 3):

$$DV \sim Setting \times MCI + Start\_Order + (1 | Part\_ID) + (1 | Dyad\_ID)$$

**Equation 3**. *Linear mixed-effects model specification including Setting × MCI interaction, Start Order, and random intercepts for participant (Part_ID) and dyad (Dyad_ID).*

## Qualitative Analysis

Qualitative data were analyzed using thematic analysis (Braun & Clarke, 2006), combining inductive coding with theory-guided interpretation (CLT, EPT, SDT). Coding progressed from semantic codes to themes. Trustworthiness procedures included auditable coding, retention of contradictory cases, and quote-linked evidence.

## Mixed-Methods Integration

Quantitative and qualitative findings were integrated using joint-display logic to assess convergence, complementarity, divergence, and expansion (Fetters et al., 2013; Guetterman et al., 2015).

## Results

### Task Duration

Task duration was substantially shorter in the Open shelving condition ($M = 314.06$ s, $SD = 91.73$) than in the Closed condition ($M = 617.53$ s, $SD = 172.98$; see Figure 3a). The linear mixed-effects model confirmed a significant main effect of Setting ($\beta = -291.20$, $SE = 52.99$, $t = -5.50$, $p < .001$), indicating that open shelving reduced task completion time by nearly five minutes on average. No significant main effect of MCI group or Setting × MCI interaction was observed.

### Physical Activity Level

Physical activity level (ENMO mean) was lower in the Open condition ($M = 0.020$, $SD = 0.005$) than in the Closed condition ($M = 0.024$, $SD = 0.007$; see Figure 3b), reflecting fewer and less extensive movements during item retrieval. The mixed-effects model confirmed a significant main effect of Setting ($\beta = -0.00615$, $SE = 0.00200$, $t = -3.07$, $p = .008$). No significant main effect of MCI group or Setting × MCI interaction was observed.

### Perceived Exertion (Borg)

Perceived exertion measured using the Borg scale was lower in the Open condition ($M = 7.76$, $SD = 2.41$) than in the Closed condition ($M = 8.47$, $SD = 2.45$). However, this difference did not reach statistical significance ($\beta = -1.50$, $SE = 0.809$, $t = -1.85$, $p = .084$). No significant main effect of MCI group or Setting × MCI interaction was observed.

### Cognitive Load

#### Objective Cognitive Load (Gaze Entropy)

Gaze entropy was higher in the Open condition ($M = 3.88$, $SD = 1.11$) than in the Closed condition ($M = 3.15$, $SD = 1.01$; see Figure 3d), indicating more distributed visual scanning under open shelving. The mixed-effects model confirmed a significant main effect of Setting ($\beta = 1.29$, $SE = 0.315$, $t = 4.10$, $p = .001$) and a significant Setting × MCI interaction ($\beta = -1.54$, $SE = 0.502$, $t = -3.07$, $p = .009$; see Figure 4), indicating that the entropy increase was larger for the non-MCI group.

#### Subjective Cognitive Load (NASA-TLX)

Perceived workload (NASA-TLX) was descriptively lower in the Open condition ($M = 6.61$, $SD = 3.49$) than in the Closed condition ($M = 8.32$, $SD = 3.16$; see Figure 3c), but this difference did not reach statistical significance ($\beta = -1.28$, $SE = 0.780$, $t = -1.65$, $p = .121$). No significant main effect of MCI group or Setting × MCI interaction was observed.

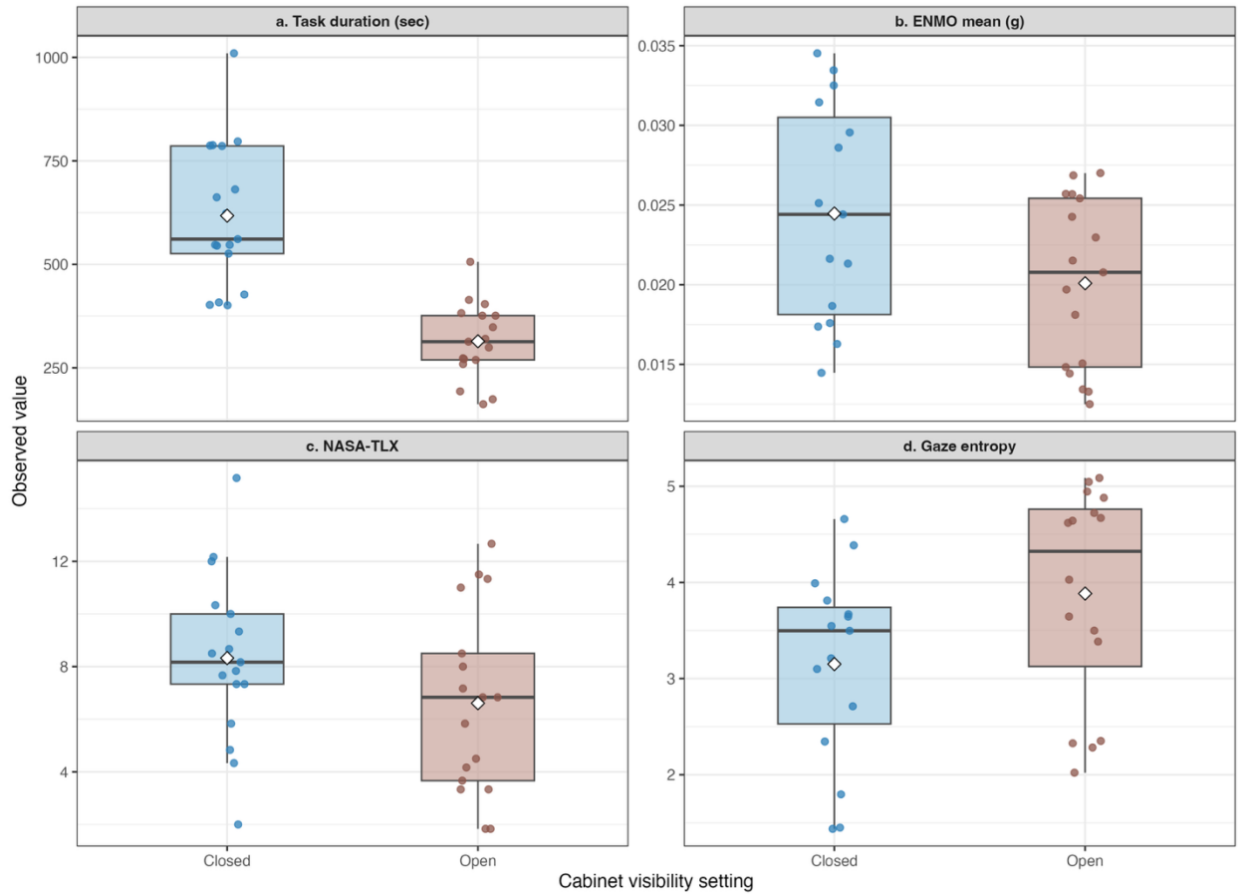

*Figure 3.* Primary outcomes by cabinet visibility condition. Observed distributions for (a) task duration, (b) ENMO mean, (c) NASA-TLX, and (d) gaze entropy in the Closed and Open shelving conditions. Points show individual observations, boxplots show distribution summaries, and white diamonds indicate condition means.

**Motivation**

Intrinsic motivation showed small descriptive differences between conditions. Interest/enjoyment was slightly higher in the Open condition ($M = 5.96$, $SD = 0.94$) than the Closed condition ($M = 5.75$, $SD = 0.81$); perceived competence was higher in Open ($M = 5.89$, $SD = 0.96$) than Closed ($M = 5.31$, $SD = 1.38$); and pressure/tension was lower in Open ($M = 2.58$, $SD = 1.54$) than Closed ($M = 2.88$, $SD = 1.29$). None of these differences were statistically significant (interest/enjoyment: $\beta = 0.00$, $SE = 0.191$, $p = 1.000$; perceived competence: $\beta = 0.68$, $SE = 0.401$, $p = .111$; pressure/tension: $\beta = -0.28$, $SE = 0.374$, $p = .466$). No significant main effects of MCI group or Setting × MCI interactions were observed.

**Exploratory Analyses**

MoCA moderated gaze entropy, with a significant Setting × MoCA interaction ($\beta = 0.932$, $SE = 0.203$, $t = 4.59$, $p < .001$), indicating greater increases in gaze entropy in the Open condition at higher MoCA scores.

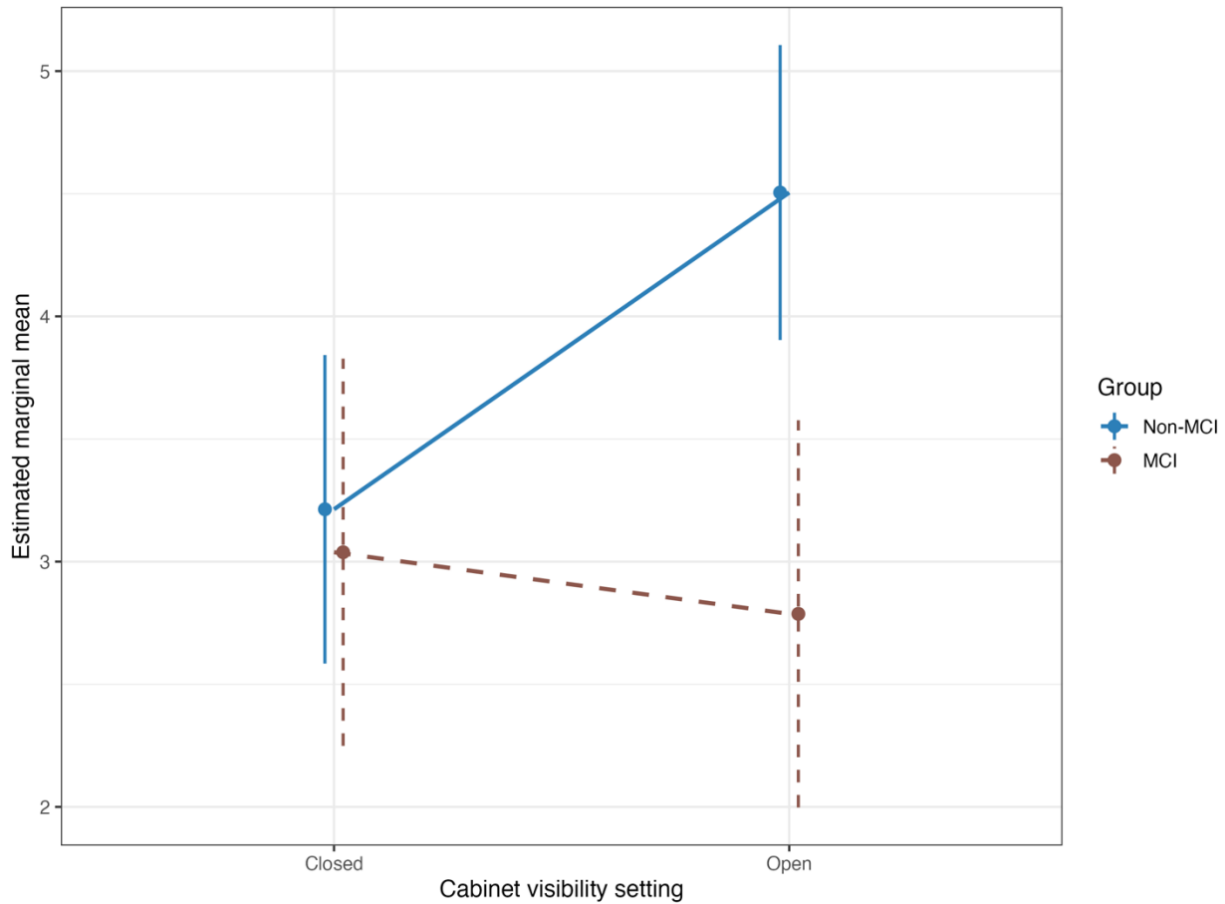

*Figure 4.* Setting by group interaction for gaze entropy. Estimated marginal means and 95% confidence intervals for gaze entropy in the Closed and Open shelving conditions for participants with MCI and without MCI. The blue solid line represents the non-MCI group, while the red dashed line represents the MCI group.

No significant moderation effects were observed for task duration (p = .892), ENMO mean (p = .387), NASA-TLX (p = .248), Borg ratings (p = .265), or fixation-related measures (p = .979).

Spearman correlations showed weak, non-significant associations between NASA-TLX and objective measures overall. Within the Open condition, Borg ratings were positively associated with ENMO mean (ρ = .484, p = .049, n = 17).

**Robustness Checks**

Sensitivity analyses showed that main effects of cabinet visibility on task duration, physical activity level, and gaze entropy remained stable across random-effects specifications.

Leave-one-participant-out analyses and internal prediction checks showed weak cross-condition prediction (e.g., task duration r = −0.06), indicating condition-specific behavior rather than stable individual traits.

**Qualitative Findings**

Thematic analysis of interview data identified several recurring themes related to visual accessibility and task performance. Where relevant, themes are discussed in terms of differences between participants with MCI and those without MCI.

**Theme 1: Visual Accessibility as Cognitive Offloading**

Participants consistently described open shelving as supporting faster item identification and retrieval, noting that visibility reduced the need for memory-based search. One participant explained, *"You can see where everything is. You don't have to guess,"* while another stated that the task required *"less memory… less thinking I have to remember where it is."* These perceptions were particularly emphasized among participants with MCI.

**Theme 2: Effort–Independence Paradox**

Despite improved efficiency in the Open shelving condition, some participants associated Closed cabinets with greater independence. One participant noted, *"The closed cabinet would make me feel more independent,"* while another stated that retrieving items from Closed cabinets felt more rewarding because *"I had to work for it."* This highlights a distinction between objective efficiency and perceived competence.

**Theme 3: Perception–Adoption Gap**

Participants often acknowledged the functional benefits of open shelving but expressed reluctance to adopt it in their own homes. Concerns centered on aesthetics and organization, with comments such as *"It would look messy"* and *"everybody doesn't need to know what you have."* Issues of dust and maintenance were also raised.

**Theme 4: Safety as a Latent Factor**

Safety concerns arose regarding cabinet interaction. One participant described, *"I have opened a cabinet… stood up, and banged my head… safety first,"* while others noted risks associated with open cabinet doors and awkward reach positions.

**Theme 5: Habituation and Spatial Memory**

Some participants indicated that familiarity with their home kitchen reduced the perceived benefits of open shelving. As one participant stated, *"I know exactly where everything is,"* suggesting that established spatial memory can mitigate search difficulty in Closed cabinet environments. This pattern was notably more pronounced among non-MCI participants; those with MCI reported greater uncertainty about item locations, suggesting that open shelving may offer a greater advantage for individuals whose spatial memory is less reliable.

**Theme 6: Care-Partner Endorsement**

Care partners consistently endorsed open shelving as beneficial for individuals with MCI. One noted, *"Because he forgets where things are,"* while another explained that visible storage would help because

*"he would be able to more easily see where things are."* However, they also acknowledged practical concerns related to organization and maintenance.

**Mixed-Methods Integration**

Integration of quantitative and qualitative findings revealed convergent and complementary patterns, as summarized in Table 3. Objective measures showed that open shelving reduced task duration and physical activity level while altering visual attention patterns (gaze entropy). Qualitative findings from the semi-structured interviews supported these results by highlighting reduced search effort and increased ease of item retrieval in the Open shelving condition.

The divergence between subjective and objective measures of cognitive load was also reflected across data sources. While the subjective workload measure (NASA-TLX) did not show significant differences between conditions, qualitative interview responses suggested perceived reductions in effort and increased ease of task completion under open shelving.

Qualitative findings also extended beyond objective outcome measures, identifying themes related to adoption preferences, safety considerations, and perceptions of independence, highlighting additional factors that influence the real-world implementation of visually accessible kitchen designs.

*Table 3. Joint display linking quantitative outcomes with qualitative themes.*

| Integrated domain | Quantitative evidence | Qualitative evidence | Integration judgment | Meta-inference |
|---|---|---|---|---|
| Task efficiency | Open reduced task duration (b = −291.20, p < .001) | Repeated reports of faster findability and less searching in Open | Convergence | Open likely improves efficiency through reduced search latency |
| Physical activity burden | Open reduced ENMO mean (b = −0.00615, p = .008) | Reports of fewer unnecessary movements and less door handling | Convergence | Reduced movement burden is both measurable and experientially visible |
| Eye-tracking indicators of cognitive load (gaze entropy) | Open increased gaze entropy (b = 1.29, p = .001); Setting × MCI significant (p = .009) | Reports of direct retrieval and reduced guessing, with stronger urgency in MCI-linked narratives | Convergence with group qualification | Open changes scanning strategy, and group status moderates magnitude |
| Global workload and exertion | NASA-TLX and Borg setting effects were non-significant | Interviews described local micro-frictions and local ease shifts | Complementarity | Broad scales and interview narratives capture different resolution levels |
| Independence and competence | Objective efficiency favored Open | Subset linked Closed to mastery and agency | Partial divergence | Efficiency and perceived mastery are distinct constructs, consistent with SDT |
| Home adoption intention | Not a primary quantitative endpoint | Broad reluctance to full home Open adoption due to social-aesthetic concerns | Expansion beyond quantified scope | Efficacy and adoption are distinct implementation targets |
| Safety and accessibility | Not directly captured in VR endpoint set | Head-strike and reach-risk narratives emerged clearly | Expansion beyond quantified scope | Interviews identify real-world constraints for translation |
| Care-partner perspective | No direct quantitative proxy | All care partners endorsed Open for recipient benefit with caveats | Triangulating support | Third-party observations reinforce functional benefit signal |

## Discussion

This study examined how visual accessibility through open shelving influences task performance, cognitive load, physical activity level, intrinsic motivation, and user experience in a virtual kitchen among older adults with and without MCI. Open shelving significantly reduced task duration and physical activity level and altered gaze-based visual search behavior, while subjective workload, motivation, and categorical MCI group differences were not significant.

The reduction in task duration observed in the Open shelving condition indicates that visual accessibility improves task efficiency by reducing search-related demands. This interpretation is supported by qualitative findings, where participants consistently described the Open shelving condition as making items easier to find and reducing the need to guess or remember locations. These findings are consistent with prior work showing that less visible storage increases task difficulty and search time in older adults (Machry et al., 2025) and align with findings from Bilau et al. (2025), which also demonstrated reduced task duration under open shelving in a physical kitchen environment.

The reduction in physical activity level further supports this interpretation. Lower ENMO values in the Open shelving condition indicate that participants required fewer movements to complete the task, reflecting reduced exploratory behavior and fewer cabinet interactions. Similar reductions in movement demands under open shelving have been reported in Bilau et al. (2025), suggesting that visual accessibility consistently reduces physical effort across both physical and virtual environments. From the perspective of EPT, these findings indicate that increasing environmental transparency reduces environmental demands, enabling more efficient interaction with the environment (Lawton & Nahemow, 1973).

A key finding of this study is the divergence between subjective and objective indicators of cognitive load. NASA-TLX scores did not differ significantly across cabinet conditions, whereas gaze entropy increased in the Open shelving condition and was moderated by cognitive status. This pattern suggests that visual accessibility altered visual-search organization without being reflected in global subjective workload ratings. From the perspective of CLT, this indicates that environmental design can influence cognitive processing at a level not fully captured by self-report measures (Sweller, 1988; Paas et al., 2003). The Open shelving condition likely reduced extraneous cognitive load associated with search and recall by enabling recognition-based retrieval, but these changes may not have been consciously perceived as differences in overall workload during a short-duration task. A similar divergence between subjective and objective measures was observed in Bilau et al. (2025).

The moderation effect observed for gaze entropy further suggests that cognitive functioning operates along a continuum rather than discrete categories. Higher MoCA scores were associated with greater increases in gaze entropy in the Open shelving condition, indicating more distributed visual-search patterns when items were visible. In contrast, participants with lower cognitive scores showed less variation across conditions. This pattern suggests that individuals with higher cognitive capacity may be better able to take advantage of visually accessible environments, while those with cognitive impairment may rely on more constrained search strategies (Cabeza et al., 2018; Salthouse, 2012).

Intrinsic motivation did not differ significantly across cabinet conditions, despite clear differences in objective task performance. Qualitative findings provide additional context for this result. While participants recognized that the Open shelving condition made tasks easier, some associated the Closed cabinet condition with greater independence and effort. For these participants, successfully recalling item locations was linked to a sense of competence and mastery. This pattern reflects the distinction between efficiency and perceived competence described in SDT (Deci & Ryan, 2000). Similar patterns were observed in Bilau et al. (2025), where participants acknowledged functional benefits of open shelving while expressing preferences shaped by familiarity and perceived control.

The pattern of convergence and divergence across data sources has theoretical implications. The alignment between reduced task duration and qualitative reports of faster retrieval supports the interpretation that visual accessibility reduces extraneous search demands in a manner that is both objectively measurable and perceived by participants. The absence of a corresponding effect on NASA-TLX scores, however, suggests that global subjective workload measures may not be sensitive to design-level changes in extraneous cognitive load, particularly in short-duration tasks where overall effort perceptions are dominated by intrinsic task demands. The divergence between objective task efficiency and perceived competence in the closed cabinet condition further indicates that functional performance and subjective agency are distinct constructs, a distinction relevant for design acceptance beyond functional benefit.

These findings suggest that visual accessibility can serve as an effective design strategy for improving functional performance in kitchen environments by reducing search demands and movement burden. However, adoption of visually accessible designs may depend on contextual factors. Participants expressed concerns related to clutter, organization, and visibility of stored items, indicating that fully open shelving may not be acceptable in all home environments. Safety considerations, including cabinet-door contact and reach-related strain, further highlight the need to consider ergonomic risks in design. These findings suggest that hybrid or adaptive solutions that balance visibility with enclosure may be more appropriate for real-world implementation, consistent with user-centered design approaches (Czaja et al., 2019).

Several limitations should be considered when interpreting these findings. The sample size was relatively small, which may limit generalizability and reduce statistical power for detecting interaction effects. The study was conducted in a virtual reality environment, which, although designed to replicate a real-world kitchen, may not fully capture long-term behavior or real-world use patterns. The tasks were short in duration, and cumulative benefits in task performance, cognitive load, intrinsic motivation, and user experience over time were not assessed. In addition, subjective measures such as NASA-TLX and IMI may not fully capture underlying cognitive and motivational processes, as reflected in the divergence between subjective and objective findings.

Future research should extend these findings by examining visual accessibility in more ecologically valid contexts, including studies conducted in participants' home environments. In addition, qualitative findings highlight the importance of involving users in the design process to address concerns related to aesthetics, organization, and safety. Participatory and co-design approaches may therefore be valuable for developing adaptive kitchen cabinet systems that balance visibility with enclosure and align with user preferences and real-world constraints.

## Conclusion

Visual accessibility through open shelving improved task efficiency and reduced physical activity level during kitchen tasks, while altering gaze-based indicators of visual-search organization. However, these improvements were not reflected in subjective workload or intrinsic motivation, highlighting a divergence between objective and subjective measures. The findings demonstrate that environmental design can meaningfully shape task performance and cognitive processing, with effects moderated by continuous cognitive status. These results provide empirical support for incorporating visual accessibility into cognitively supportive kitchen design while emphasizing the need for solutions that balance usability with user preferences, safety, and real-world adoption considerations.

## Acknowledgment

We thank Bolaji Omofojoye, Elizabeth Vanderburg, and Sonya Williams for their invaluable contributions to this study. We also thank the members and care partners at the Charlie and Harriet Shaffer Cognitive Empowerment Program, Campbell-Stone Senior Living, and A.G. Rhodes Nursing Home for participating in this study, and their staff for supporting our research.

## Key Points

- Visual accessibility through open shelving significantly reduced task duration and physical activity level during a virtual kitchen task.
- Gaze entropy increased under open shelving, indicating changes in visual-search organization, with effects moderated by continuous cognitive status.
- Subjective workload (NASA-TLX) and intrinsic motivation (IMI) did not differ significantly between cabinet conditions.
- Qualitative findings revealed reduced search effort alongside concerns related to aesthetics, safety, and real-world adoption.
- Findings highlight a divergence between objective performance and perceived experience and support visually accessible, adaptive kitchen design for aging populations.